\documentclass[a4paper,10pt]{article}
\newcommand{\head}[1]{\textnormal{\textbf{#1}}}
\usepackage[onehalfspacing]{setspace}
\usepackage{amsmath}
\usepackage{amssymb}
\usepackage{hyperref}
\usepackage{fontenc}
\usepackage{inputenc}
\usepackage{authblk}
\usepackage{color}
\usepackage{graphicx}
\usepackage{cite}
\usepackage{slashed}
\hypersetup{colorlinks=true,linkcolor=magenta,linktoc=page,citecolor=blue}
\usepackage{tikz}
\usepackage{multirow}
\usepackage{hhline}
\usepackage[top=1.5cm,bottom=2.5cm,right=2.54cm,left=2.54cm]{geometry}
\definecolor{c1}{rgb}{0.5,0,1}
\colorlet{aqua}{-red!75}
\newcommand{\be}{\begin{equation}}
\newcommand{\bea}{\begin{eqnarray}}
\newcommand{\eea}{\end{eqnarray}}
\newcommand{\ba}{\begin{array}}
\newcommand{\ea}{\end{array}}
\newcommand{\ee}{\end{equation}}
\numberwithin{equation}{section}

\begin{document}
\begin{flushright}
 IPM/P-2015/071 \\
\end{flushright}

\vspace*{20mm}
\begin{center}
{\Large {\bf  Thermalization of Mutual Information in Hyperscaling
Violating Backgrounds}\\}

\vspace*{15mm} \vspace*{1mm} {M. Reza Tanhayi}

 \vspace*{1cm}

{Department of Physics, Faculty of Basic Science, Islamic Azad
University Central Tehran Branch (IAUCTB), P.O. Box 14676-86831,
Tehran, Iran\\
School of Physics, Institute for Research in Fundamental Sciences
(IPM) P.O. Box 19395-5531, Tehran, Iran}

 \vspace*{0.5cm}
{E-mail: {\tt mtanhayi@ipm.ir}}

\vspace*{1cm}
\end{center}

\begin{abstract}

We study certain features of scaling behaviors of the mutual
information during a process of thermalization, more precisely we
extend the time scaling behavior of mutual information which has
been discussed in \cite{Alishahiha:2014jxa} to time-dependent
hyperscaling violating geometries.  We use the holographic
description of entanglement entropy for two disjoint system
consisting of two parallel strips whose widths are much larger
than the separation between them. We show that during the thermalization
process, the dynamical exponent plays a crucial rule in reading the general time scaling behavior of mutual information
(\emph{e.g.,} at the pre-local-equilibration regime). It is shown that the scaling violating parameter can be employed to define an effective dimension.

\end{abstract}
\newpage

\section{Introduction}

The AdS/CFT correspondence, sometimes called as gauge/gravity duality, is a conjectured relationship between quantum field theory and gravity. Precisely, in this correspondence, quantum physics of strongly correlated many-body systems is related to the classical dynamics of gravity which lives in one higher dimension.
On the other hand according to the AdS/CFT dictionary, an AdS
geometry at the gravity side could only address the conformal
symmetry of the dual field theory. However, the generalization of gauge/gravity correspondence to geometries which are not
asymptotically AdS seems to be important as long as such extension
may be related to the invariance under a certain scaling of dual
field theory which does not even have conformal symmetry. Such
generalization of AdS is actually motivated by consideration of
gravity toy models in condensed matter physics (the application of
such generalization can be found, for example, in
\cite{Hartnoll:2009sz}). A prototype of this generalization is a
theory with the Lifshitz fixed point in which the spatial and time
coordinates of a field theory have been scaled as \be\label{scale}
t\rightarrow \zeta^z
t,\,\,\,\,\,\,\,\,\,\,\,\,\,\vec{x}\rightarrow\zeta\vec{x},\,\,\,\,\,\,\,\,\,\,\,\,\,r\rightarrow
\zeta r,\ee where $z$ is the critical dynamical exponent. From the
holographic duality point of view, for a $(D + 1)$-dimensional
theory, the corresponding $(D + 2)$-dimensional gravity dual can
be defined by the following metric \be \label{lif} ds^2_{D+2} =
\frac{-dt^2}{r^{2z}} +\frac{dr^2}{r^2}
+\frac{1}{r^2}\sum_{i=1}^Ddx_i^2,\ee where in this paper the AdS
radius is set to be one. Due to the anisotropy between space and
time, it is clear that this metric can not be an ordinary solution
of the Einstein equation, in fact one needs some sorts of matter
fields to break the isotropy, e.g., by adding a massive vector
field or a gauge field coupled to a scalar field
\cite{Kachru:2008yh, Balasubramanian:2008dm, Alishahiha:2012qu,
Roychowdhury:2015fxf}. In general by adding a dilaton with non
trivial potential and an abelian gauge field to Einstein-Hilbert
action (Einstein-Maxwell-Dilaton theory), one can find even more
interesting metrics, in particular the following metric has been
used frequently \cite{Charmousis:2010zz} \be\label{lif2}
ds^2_{D+2} =r^{\frac{2\theta}{D}} \Big(\frac{-dt^2}{r^{2z}}
+\frac{dr^2}{r^2} +\frac{1}{r^2}\sum_{i=1}^Ddx_i^2\Big),\ee where
$\theta$ is hyperscaling violation exponent. This metric under the
scale-transformation \eqref{scale} transforms as
$ds\rightarrow\zeta^{\frac{\theta}{D}}ds$. In a theory with
hyperscaling violation, the thermodynamic parameters behave in
such a way that they are stated in $D-\theta$ dimensions; More
precisely, in a $(D + 1)$-dimensional theories which are dual to
background \eqref{lif}, the entropy scales with temperature as
$T^{D/z}$, however, in the presence of $\theta$ namely dual to
\eqref{lif2}, it scales as $T^{(D-\theta)/z}$
\cite{Charmousis:2010zz, Gouteraux:2011ce}. Therefore one may
associate an effective dimension to the theory and this becomes
important in studying the log behavior of the entanglement entropy
of system with Fermi surface in condense matter physics,
explicitly it was shown that for $\theta = D-1$ for any $z$, the
entanglement entropy exhibits a logarithmic violation of the area
law \cite{Alishahiha:2012cm}. On the other hand for such
backgrounds time-dependency can be achieved by Vaidya metric with
a hyperscaling violating factor. It is the main aim of this paper to investigate how entanglement (mutual information) spreads in time-dependent hyperscaling violating backgrounds.

Basically, the AdS-Vaidya metric is used to describe a
gravitational collapse of a thin shell of matter in formation of
the black hole. This metric in $D+2$ dimensions is given by
\be\label{hyper}
 ds^2=\frac{1}{\rho^2}\bigg(-f(\rho,v)dv^2-2 d\rho dv+\sum_{i=1}^{D}dx_i^2\bigg),\;\;\;\;\;\;\;\;\;
f(\rho,v)=1-m(v)\rho^{D+1}, \ee where $\rho$ is the radial
coordinate, $x_i$'s $(i=1,..,D)$ are spatial boundary coordinates
and, here, the mass $m(v)$ is supposed to be an arbitrary
function of the null coordinate\footnote{ We should mention that
the mass function satisfies some boundary condition in time, e.g.
for the initial time it is zero and for the late time it is a
constant: transition from pure AdS to AdS black hole.} $v$.
Holographic dual of such background can be described by a system which
undergoes a sudden change which might be caused by turning on a uniform density of
sources for a short time interval $\delta t$ at $t = 0$ and then turning it
off. This process is called as a quantum quench which can excite the system to an
excited state with non-zero energy density. Evolution towards an equilibrium state after a global quantum quench \cite{CC} is an example of
the thermalization. During thermalization (being out of
equilibrium) the usual thermodynamical quantities such as thermal
entropy and pressure are not well-defined quantities. So that such
an evolution cannot be studied within the context of standard
thermodynamics, however, the entanglement entropy and the mutual
information could potentially play a key role in probing the
process of thermalization in those systems. Based on AdS/CFT correspondence, a global quench in the boundary theory can be described
by a thin shell of matter starting from the boundary and
collapsing to form a black hole and the corresponding metric is given by \eqref{hyper}. On the other hand, the covariant holographic entanglement entropy proposal \cite{Dong:2012se} suggests that entanglement entropy is proportional to the extremal surface in the bulk where this surface anchors on the boundary of entangling region on the conformal boundary of the bulk. This leads to the fact that entanglement entropy of the boundary region, will depend on time. In the process of thermalization, the local equilibrium which scaled by the horizon radius plays a role of defining a time scale when the system has
ceased production of thermodynamic entropy though the entanglement entropy still increases \cite{Liu:2013iza}.

In hyperscaling violating time-dependent geometries entanglement
entropy has been studied in \cite{{Alishahiha:2014cwa},{Fonda:2014ula}}. In this work we study
the time scaling behavior of the mutual information for a system
after a global quantum quench. To do so we use Vaidya metric with
hyperscaling violating factor, and the main aim of the present
study is to extend the consideration of \cite{Alishahiha:2014jxa}
in studying different scaling behaviors of the mutual information
after a global quantum quench.

The organization of the paper is as follows. In the next section
we will study the computations of holographic mutual information
in a hyperscaling violating background. In section three we will
study the scaling behaviors of the mutual information during the
thermalization process. Last section is devoted to
discussions. In the appendix we will review some related
computations of the entanglement entropy.


\section{Mutual Information in Hyperscaling Violating Backgrounds}

Entanglement entropy is a measure of storing quantum information
in a quantum state. It is indeed a remarkable tool in studying
quantum systems which has been deduced from the first principles
of quantum mechanics. Beside the entanglement entropy, for two
disjoint regions say as $A$ and $B$, when we are interested in the
amount of information that these two systems could share, the
mutual information is mostly used. The mutual information can be
expressed in terms of the entanglement entropy as \be\label{mutu}
I(A,B)=S(A)+S(B)-S(A\cup B), \ee where $S(A),\, S(B)$ and $S(A\cup B)$
are  entanglement entropies of the regions $A$, $B$ and their
union, respectively with the rest of the system. Entanglement
entropy for two disjoint regions has been studied in
\cite{{Caraglio:2008pk},{Furukawa:2008uk},{Calabrese:2009ez},{Calabrese:2010he}}\footnote{
To study the holographic entanglement entropy of multiple strips
in various holographic theories see also
Ref.\cite{Ben-Ami:2014gsa}}, therefore, one can use the results to
study the mutual information. From the definition of mutual
information and after making use of the subadditivity property of
the entanglement entropy, it is evident that in computing the
mutual information there is no UV divergency and it is always
non-negative and  it becomes zero if two system are uncorrelated.
On the other hand from the AdS/CFT correspondence, one can show
that there is a phase transition from zero to a positive value in
the mutual information as one decreases the distance between two
systems \cite{{Headrick:2010zt},{Hubeny:2007re},{Tonni:2010pv}}.
In Ref.\cite{Alishahiha:2014jxa}, for a certain entangling region
namely for two strips, the various scaling behaviors of the mutual
information during a process of thermalization after a global
quantum quench have been studied. Here, in this paper we extend the
results in parts to study the scaling behaviors of the mutual
information in hyperscaling violating backgrounds. From the
gauge/gravity duality the thermalization process after a global
quantum quench is equivalent to the black hole formation at the
gravity side due to a gravitational collapse. The corresponding
metric is given by \eqref{hyper}. For example entanglement entropy
has been studied in
\cite{{AbajoArrastia:2010yt},{Balasubramanian:2010ce},{Aparicio:2011zy},{Galante:2012pv},{Caceres:2012em},{Baron:2012fv},
{Fischler:2012ca},{Fischler:2012uv},{Caputa:2013eka},{Fischler:2013fba}}
and mutual information in
\cite{{Balasubramanian:2011at},{Allais:2011ys},{Callan:2012ip},{Li:2013sia}}.
Now it is natural to study the time-dependency of the holographic
entanglement entropy and mutual information in time-dependent
hyperscaling violating background.

As mentioned, the hyperscaling violating geometries can be
obtained by adding a scalar field with nontrivial potential
(dilaton) and an abelian gauge field to the Einstein-Hilbert
action. The general metric is given by \be\label{met1}
ds^2_{D+2}=r^{-2\frac{\theta}{D}}\Bigg(-r^{2z}f(r)dt^2+\frac{dr^2}{r^2f(r)}+r^2\sum_{i=1}^Ddx_i^2\Bigg),
\ee where $z$ and $\theta$ are the dynamical and hyperscaling
violation exponents, respectively and \be 
f(r)=1-\frac{m}{r^{D-\theta+z}}. \ee The above metric belongs to a
neutral black brane solution with horizon radius $r_H$.\footnote{
$f(r_H)=0$ gives us the horizon radius.}  The important fact is
that in the present model three free parameters appear: dimension
of the space-time $D$, scaling violating parameter $\theta$ and
dynamical exponent $z$. However, it is shown that the entanglement
entropy and the mutual information up to an overall factor of
$L^{D-1}$ are sensitive to the dynamical exponent $z$ and the
effective dimension $d$ which is defined by \be \label{effec}
d\equiv D-\theta+1,\ee the overall factor can be fixed by a
dimensional analysis. Moreover, in Ref. \cite{Dong:2012se}, the authors have shown that the null energy condition at the gravity side imposes the following constrains on the free parameters
\begin{eqnarray}
(D-\theta)(z-1-\frac{\theta}{D})\ge0,\hspace{1cm}(z-1)(D+z-\theta)\ge0,
\end{eqnarray}   
in this paper we will consider $z>1$ and $D>\theta$.\footnote{We thank the referee for his/her comment on this point.}  

In what follows in this section, we will compute the holographic
mutual information for two parallel infinite strips with the equal
width $\ell$ separated by a distance $h$ in a $D+1$ dimensional
field theory in both static and time-dependent hyperscaling
violating backgrounds. We will also review the computation for
entanglement entropy of a strip in such backgrounds in Appendix A.

In Ref. \cite{Headrick:2010zt} it was shown that there would be a
first order phase transition in computing the holographic mutual
information when the distance between two strips increases. More
precisely the value of $\frac{h}{\ell}$ plays a crucial role in
which above it the mutual information vanishes. This fact comes
form the definition of entanglement entropy of the union $A\cup B$
and for a given two strips with the widths $\ell$ and distance
$h$, from the holographic assumption point of view, this phase
transition may be understood from the fact that, there are two
minimal hypersurfaces associated with the entanglement entropy
$S(A\cup B)$, consequently, the corresponding entanglement entropy
behaves differently. This means that for two cases $h\ll \ell$ and
$h\gg \ell$ one gets \bea\label{SAUB} S({A\cup B})=\Bigg\{
\begin{array}{rcl}
&S(2\ell+h)+S(h)&\,\,\,h\ll \ell,\\
&2S(\ell)&\,\,\,h\gg \ell,
\end{array}\,\,
\eea where $S(l)$ stands for the entanglement entropy of a strip
with width $l$. Making use the above relation and also
\eqref{mutu}, clearly the mutual information becomes zero in the
case of $h\gg \ell$, on the other hand for $h\ll \ell$, one finds
\be\label{MI} I(\ell,\ell,h)=2S(\ell)-S(2\ell+h)-S(h)\equiv I. \ee
In what follows we will also consider $h\ll \ell$. This indicates
that in order to find the mutual information of two parallel
strips, the entanglement entropy of three strips should be
computed with widths $h, \ell$ and $2\ell+h$. We emphasize that in
this paper we consider the case in which the widths of two
parallel strips are same, otherwise one has to compute four
entanglement entropies corresponding to $\ell_1, \ell_2, h $ and
$\ell_1+\ell_2+h$.

First let us write the mutual information for two parallel strips
in the vacuum state, it is indeed achieved from the entanglement
entropy for a strip in a $D+1$-dimensional CFT whose gravity dual
is provided by the hyperscaling violating geometry. The corresponding entanglement
entropies are given by \eqref{SV}, therefore one obtains
\be\label{vac}
{I}_{\text{vac}}=\frac{L^{D-1}c_{d-1}}{4G_N}\left(-\frac{2}{\ell^{d-2}}+\frac{1}{(2\ell+h)^{d-2}}+\frac{1}{h^{d-2}}\right),
\ee where $d$ is the effective dimension and $c_{d-1}$ is a
coefficient are given by \eqref{effec} and \eqref{cdef},
respectively. The condition $h\ll \ell$ guarantees the positivity
of resultant mutual information. Note that as one increases
the width of strips there is an upper limit for the mutual
information given by \be\label{IMaxv} {I}_{\text{vac}}^{\rm
max}=\frac{L^{D-1}}{4G_N}\;\frac{c_{d-1}}{h^{d-2}}, \ee which is
the absolute value of the finite part of the entanglement entropy
for a strip with the width $h$.

On the other hand in order to compute the mutual information of
the same strips in thermal states one can use the corresponding
gravity dual which is provided by an AdS black brane hyperscaling
violating metric \eqref{met1}. In general for the entanglement
entropies there is no an analytic expression but in some certain
limits, e.g., $\ell\ll \rho_H$ one can compute the entanglement
entropy which is reviewed in appendix A. In this limit noting that
since $h\ll \ell$ one also has $h\ll \rho_H$, all the entanglement
entropies involved  in the computation of the mutual information,
may be expanded as equation \eqref{SATH} and consequently the
corresponding mutual information in this background is given by
\bea\label{IBH} {I}_{\text{BH}}={I}_{\text{vac}}
-\frac{L^{D-1}c^z_1}{16G_N(d-2)}\; \frac{
(\ell+h)^{1+z}}{\rho_H^{d-1+z}}, \eea note that $c_1$ is given by
\eqref{c1z}. From \eqref{IBH}, one can say that for two static regions, mutual information
is maximal when the system is in the vacuum state, namely $I_{\text{vac}}> I_{\text{BH}}$.\footnote{Note that in the process of thermalization after a quench this statement should be modified, mutual information undergoes a growing and then decreasing regimes, which will be discussed in the next section. We should mention that in this paper we only consider the excited states which are thermal, but when they are created by acting some local or non-local operators on the ground state, the answer is not so clear, for more details see \cite{Fischler:2012ca}.}

Now let us consider the following case: $h\ll \rho_H \ll \ell$.
One can say that the corresponding entanglement entropy for the
region $h$ should be approximated by equation \eqref{SATH}, while
for those of $\ell$ and $2\ell+h$ one has to use the large
entangling region expansion given by equation \eqref{EEBHh},
consequently the mutual information reads as \be\label{MIS}
I=\frac{L^{D-1}}{4G_N}\left(\frac{c_{d-1}}{h^{d-2}}-\frac{c_2}{\rho_{H}^{d-2}}-\frac{h}{2\rho_H^{d-1}}-\frac{c^z_1}{4(d-2)\rho_H^{d-1+z}}h^{z+1}\right),
\ee where $c_2$ is a positive number which for some specific value
of $z$ and $d$ it is calculated numerically in
table \eqref{table}. Note that in Ref.\cite{Alishahiha:2014jxa},
it was shown that although the term containing $c_2$ is subleading
in the expression of the entanglement entropy at large entangling
region, but this term becomes significant in the expression of
mutual information. Finally for $\rho_H \ll h$ and $\rho_H \ll
\ell$ the mutual information is identically zero.


\section{Quantum Quench: Time evolution of Mutual Information in
Hyperscaling Violating Backgrounds}

In this section we will study the scaling behavior of the
holographic mutual information after a global quantum quench in a
hyperscaling violating background. According to the proposal of covariant holographic entanglement entropy, time evolution of the entanglement entropy after a global quantum quench is controlled by the geometry around and inside the event horizon of the black hole. In Ref. \cite{{Liu:2013iza},{Liu:2013qca}} for large entangling region, it was shown that certain time intervals should be considered as pre-local-equilibration with quadratic time behavior, post-local-equilibration with linear growth, late-time equilibrium. To study the evolution of mutual information, for each part of the entanglement entropies of \eqref{mutu}, one should take into account these time intervals as well. In what follows we are dealing with these intervals separately. To do so as in previous section, we shall
consider two strips with widths $\ell$ separated by distance $h$
with condition $h\ll \ell$. We are interested in time evolution of
the system, consequently, the covariant proposal of the
holographic entanglement entropy of three strips with widths $ h,
\ell$ and $2\ell+h$ in the AdS-Vaidya metric \eqref{hyper} must be
considered. This means that three hypersurfaces should be studied
which each of them has crossing point and the turning
point\footnote{The crossing point refers to a point where the
hypersurface intersects the null shell whereas $\rho_{i\; t}$
stands for the turning point of the extremal hypersurface in the
bulk.} as denoted by $(\rho_{i\;c},\rho_{i\; t})$ with $i=1, 2,
3$. If one uses the entanglement entropy as a probe, the radius of
the horizon $\rho_H$, and the size of the entangling region are
used as scales to address the time evolution of the system.
However in our case beside the radius of the horizon,
$\frac{\ell}{2},\,\,\frac{h}{2}$ and $\ell+\frac{h}{2}$ play the
crucial rule. So that depending on the size of $\rho_H$ is larger
or smaller than the entangling regions and noting that $h\ll
\ell$, one may distinguish four main possibilities for the order
of scales as follows \begin{itemize} \item  $\rho_H \ll
\frac{h}{2}$, \item  $\frac{h}{2}\ll\rho_H\ll
\frac{\ell}{2}<\ell+\frac{h}{2}$, \item  $\frac{h}{2}\ll
\frac{\ell}{2}<\rho_H<\ell+\frac{h}{2}$, \item $\frac{h}{2}\ll
\frac{\ell}{2}<\ell+\frac{h}{2}\ll \rho_H$.
\end{itemize} In what
follows we will consider the cases separately.

\subsection{First regime: $\rho_H\ll \frac{h}{2}$}

In order to study the time evolution of mutual information we
should compute the co-dimension two hypersurfaces of the all
entangling regions which are associated with the entanglement
entropies appeared in equation \eqref{MI}. But in this case it is
important to note that the hypersurfaces cross the null shell and
could probe the $v<0$ region. In this case five separated time
intervals might be considered as: \bea t\ll
\rho^z_H,\hspace{2cm}\rho^z_H\ll t\ll
\rho_H^{z-1}\frac{h}{2},\hspace{2cm}\rho_H^{z-1}\frac{h}{2}\ll t
\ll \rho_H^{z-1}\frac{\ell}{2},\nonumber\\
\rho_H^{z-1}\frac{\ell}{2} < t <
\rho_H^{z-1}(\ell+\frac{h}{2}),\hspace{3cm}\mbox{Saturation time.
}\hspace{1.8cm}\eea Let us study the time evolution in each case.

\subsubsection{$t\ll \rho^z_H$}

At the very early time after the quench, all the co-dimension two
hypersurfaces in the bulk cross null shell where the crossing
points are very close to the boundary this means
$\frac{\rho_{c}}{\rho_H}\ll 1$, so that $\rho_{i\; c}$
approximately are the same and 
Therefore one can expand $\ell$ and $t$ which are given by
\eqref{earlytime} and also ${\cal A}$ in this limit and hence
one obtains \be\label{t} t\approx
\rho^z_t\int_0^{\frac{\rho_c}{\rho_t}}{d\xi}\frac{\xi^{z-1}}{f(\rho_t\xi)}=
\frac{\rho^z_c}{z}\left(1+\frac{z}{d-1+2z}\left(\frac{\rho_{c}}{\rho_H}\right)^{d-1+z}+{\cal
O}\left(\frac{\rho_{c}}{\rho_H}\right)^{2(d-1+z)}\right),\ee where
$\xi\equiv\frac{\rho}{\rho_t}$. This leads to the following
expression for the area \be\label{a} {\cal A}\approx
L^{D-1}\left[\frac{1}{(d-2)\epsilon^{d-2}}-\frac{c_{d-1}}{\ell^{d-2}}+\frac{1}{2(z+1)}\frac{\rho_c^{z+1}}{\rho_H^{d+z-1}}\Big(1+{\cal
O}\left((\frac{\rho_c}{\rho_H})^{d+z-1}\Big)\right)\right],\ee
where $\epsilon$ is the UV cutoff and $c_{d-1}$ is given by
\eqref{cdef}. Making use of \eqref{t} and \eqref{a} and
noting that $m=\rho_H^{1-d-z}$, at the leading order one finds
\be\label{leading} S(\ell_i)\approx
\frac{L^{D-1}}{4G_N}\left[\frac{1}{(d-2)\epsilon^{d-2}}-\frac{c_{d-1}}{\ell_i^{d-2}}+\frac{m}{2(z+1)}(zt)^{1+\frac{1}{z}}\right].\ee As one sees this expression is independent
of the hyperscaling violating factor $\theta$. Note that through
out this section we use a notion in which $i=1,2,3$ where
$l_1=h,\,\, l_2=\ell$ and $l_3=2\ell+h$. Plugging these
expressions into equation \eqref{MI}, one finds
\bea\label{earlytime}
I=\frac{L^{D-1}c_{d-1}}{4G_N}\left(-\frac{2}{\ell^{d-2}}+\frac{1}{h^{d-2}}+\frac{1}{(2\ell+h)^{d-2}}\right)+
{\cal O}(t^{2d})=I_{\rm vac}+{\cal O}(t^{2d}). \eea One observes
that the mutual information starts from its value in the vacuum,
$I_{\rm vac}$,  and remains fixed up to order of ${\cal
O}(t^{2d})$  at the early times. 

\subsubsection{ Steady behavior: $\rho^z_H\ll t\ll \rho_H^{z-1}\frac{h}{2}$}

In this time interval, in the case of computing the entanglement
entropy a significant observation of linear growth with time has
been observed in \cite{{Liu:2013iza},{Liu:2013qca}}, and the
extension to the hyperscaling violating backgrounds was done in
\cite{Alishahiha:2014cwa}. In Ref.\cite{Alishahiha:2014jxa} it was
argued that the mutual information inherits this linear behavior
as well. Actually there is a critical extremal surface which is
responsible for the linear growth in this time interval. More
precisely, equation \eqref{veff} might be thought of as the energy
conservation law for a one-dimensional dynamical system whose
effective action is given by $V_{eff}(\rho)$ with dynamical
variable $\rho$. Now, for $\rho_t$ being a fixed turning point,
$\rho_c$ can play as a free parameter which may be tuned to a
particular value of $\rho_c=\rho^*_c$ in a way that
\bea\label{criticalhyper} \frac{\partial
V_{eff}(\rho)}{\partial\rho}\bigg|_{\rho_m,\rho^*_c}=0,\;\;\;\;\;\;\;\;\;\;\;\;
V_{eff}(\rho)|_{\rho_m,\rho^*_c}=0, \eea where $\rho_m$ is a point
which minimizes the effective action. This indicates that if the
hypersurface intersects the null shell at the critical point, it
remains fixed at $\rho_m$. One can show that in this time interval
the main contributions to $\ell, t$ and ${\cal A}$ come from a
hypersurface which is closed to the critical extremal
hypersurface. Now we want to compute the $\ell$, $t$ and ${\cal
A}$ around the critical extremal hypersurface, to do so let us
consider $\rho_c=\rho^*_c(1-\delta)$ in which $\delta\ll 1$ when
$\rho\rightarrow \rho_m$. Assuming that both
$\frac{\rho^*_c}{\rho_t}$ and $\frac{\rho_m}{\rho_t}$ are very
smaller than one, the expression of \eqref{earlytime}  may be
approximated in the hyperscaling violating background as
\cite{Alishahiha:2014cwa} \bea t\approx
-\rho_t^z\frac{\xi_m^{2z-2}E^*}{f(\rho_t\xi_m)\sqrt{\frac{1}{2}
V''_{eff}(\rho_t\xi_m)}}\ln
\delta,\hspace{2cm}\frac{\ell}{2}\approx c
\rho_t-\frac{\rho_t}{\sqrt{\frac{1}{2}V''_{eff}(\rho_t\xi_m)}}\ln
\delta,\eea where
$V''_{eff}(\rho_t\xi_m)=\frac{\partial^2V_{eff}(\rho_t\xi_m)}{\partial\xi^2}|_{\xi_m,\xi^*_c}$
and
$c=\sqrt{\pi}\frac{\Gamma(\frac{d}{2d-2})}{\Gamma(\frac{1}{2d-2})}$,
and we have defined $E^*$ as \be\label{ees}
E^*=-(\frac{\rho_t}{\rho_m})^{z-1}\sqrt{-f(\xi_m)\left((\frac{\rho_t}{\rho_m})^{2z-2}-1\right)}.\ee
On the other hand from \eqref{an} one finds  \bea \frac{{\cal
A}_{d-1}}{L^{D-1}}&\approx&
\frac{1}{(d-2)\epsilon^{d-2}}-\frac{c_{d-1}}{\ell_i^{d-2}}+\frac{1}{\rho_t^{d+z-2}}\frac{f(\rho_t\xi_m)}{\xi_m^{2(d+z-2)}E^*}\,\,t,\\
&=&\frac{{\cal
A}_{\mbox{vac}}}{L^{D-1}}+\frac{1}{\rho_t^{d+z-2}}\frac{f(\rho_t\xi_m)}{\xi_m^{2(d+z-2)}E^*}\,\,t.\eea
Plugging these relations in \eqref{ees}, for large $\rho_t$ the
entanglement entropy reads \be\label{SBHL}
S\approx\frac{L^{D-1}}{4G_N}\left[\frac{1}{(d-2)\epsilon^{d-2}}-\frac{c_{d-1}}{\ell_i^{d-2}}+
\frac{\sqrt{-f(\rho_{m})}}{\rho_{m}^{d+z-2}}t+\cdots\right]. \ee
Note that by making use of equation \eqref{criticalhyper}, one can
obtain $\rho_m$ and $\rho^*_c$ in terms of the radius of horizon.

The mutual information is then obtained from equation \eqref{MI}
as follows \bea\label{early} I=I_{\rm vac}
+\frac{L^{D-1}}{4G_N}\left(2\frac{\sqrt{-f(\rho_{2\;m}})}{\rho_{2\;m}^{d+z-2}}-\frac{\sqrt{-f(\rho_{1\;
m}})}{\rho_{1\; m}^{d+z-2}}-\frac{\sqrt{-f(\rho_{3\;
m}})}{\rho_{3\; m}^{d+z-2}}\right)t+\cdots\, . \eea The important
fact is that since we are dealing with the large entangling
regions, the turning points of all hypersurfaces are large, one
can deduce that $\rho_{i\;m}=\rho_m$. As a result, the second term
in the above equation vanishes leading to a constant mutual
information in this time interval too. Thus starting from a static
solution one gets almost constant mutual information all the way
from $t=0$ to $t\sim \rho_H^{z-1}\frac{h}{2}$.


\subsubsection{Linear growth: $\rho_H^{z-1}\frac{h}{2}\ll t \ll \rho_H^{z-1}\frac{\ell}{2}$}

In this time interval the corresponding entanglement entropy of
$h$ saturates. The saturation time can indeed be obtained by
making use of the expression of the saturated entanglement entropy
and the intermediate entanglement entropy \eqref{SBHL}, which
one obtains the saturation time as \be\label{tsat}
t_s\sim\frac{h}{2}\rho_H^{z-1}-c_2\rho^z_H+c_{d-1}\frac{\rho_H^{d+z-2}}{h^{d-2}},\ee
where $c_{d-1}$ is given by \eqref{cdef}. So that one should use
the saturated entanglement entropy associated with $h$ which is
discussed in appendix  and given by  \be
S(h)\approx\frac{L^{D-1}}{4G_N
}\left(\frac{1}{(d-2)\epsilon^{d-2}}+\frac{h}{2\rho_H^{d-1}}
-\;\frac{c_2}{\rho_H^{d-2}}\right). \ee Whereas the entanglement
entropies associated with the entangling regions $\ell$ and
$2\ell+h$ are still increasing linearly with time which are given
by equation \eqref{SBHL}. Gathering all the results into equation
\eqref{MI} one finds \be {I}=\frac{L^{D-1}}{4G_N}\left[
-\frac{2c_{d-2}}{\ell^{d-2}}+\frac{c_{d-2}}{(2\ell+h)^{d-2}}-\frac{1}{\rho_{H}^{d-1}}\frac{h}{2}+\frac{c_2}{\rho_{H}^{d-2}}+\left(2
\frac{\sqrt{-{f}(\rho_{2\;m })}}{\rho_{2\;m}^{d+z-2}}-
\frac{\sqrt{-{f}(\rho_{3\;m
})}}{\rho_{3\;m}^{d+z-2}}\right)t+\cdots\right], \ee which can be
recast into the following form \be \label{muTu}I=I_{\rm
vac}+\frac{L^{D-1}}{4G_N}\left(\frac{c_2}{\rho_H^{d-2}}-\frac{c_{d-2}}{h^{d-2}}\right)+
\frac{L^{D-1}}{4G_N \rho_H^{d-1}} \left(v_E t-\frac{h}{2}\right),
\ee where \be v_E\equiv\rho_H^{d-1}\left(2
\frac{\sqrt{-{f}(\rho_{2\;m })}}{\rho_{2\;m}^{d+z-2}}-
\frac{\sqrt{-{f}(\rho_{3\;m })}}{\rho_{3\;m}^{d+z-2}}\right).\ee

It is worth to mention that in hyperscaling violating backgrounds
in general for $z\neq1$ the turning point in the bulk can not be
fixed only by $\rho_m$. Namely for a fixed $\rho_H$, the radial
coordinate which minimizes the effective potential is a function
of both $\rho_t$ and $\rho_c$ \cite{Alishahiha:2014cwa}. In other
words the effective potential becomes minimum at $\rho_m$ which
one finds

\be \label{1}\rho_t^{2(d-1)}=\rho_m^{2(d-1)}\frac{2\rho_m
f'(\rho_m)+(z-1)(\frac{\rho_c}{\rho_H})^{2d-2+2z}(\frac{\rho_m}{\rho_c})^{2(z-1)}}{2\rho_m
f'(\rho_m)-4(d-1)f(\rho_m)+(z-1)(\frac{\rho_c}{\rho_H})^{2d-2+2z}(\frac{\rho_m}{\rho_c})^{2(d-2+z)}}.\ee
 However, for the critical extremal hypersurface which is defined
by the conditions \eqref{criticalhyper}, imposing the condition
of critical point where at the minimum point the effective
potential is also zero, results in

\be\label{2}
\rho_t^{2(d-1)}=\rho_m^{2(d-1)}\frac{4f(\rho_m)+(\frac{\rho_c^*}{\rho_H})^{2}(\frac{\rho_m}{\rho_c})^{2(z-1)}}{4f(\rho_m)+(\frac{\rho_c^*}{\rho_H})^{2}
(\frac{\rho_m}{\rho_c})^{2(d-2+z)}},\ee this relation can indeed
fix $\rho_m$ and $\rho_c^*$, and by solving relations \eqref{1}
and \eqref{2} one can find $\rho_c^*$ and $\rho_m$. For large
entangling region (or at large $\rho_t$ limit) assuming that both
$\rho_m$ and $\rho_c^*$ remain finite one gets \be
\frac{\rho_m}{\rho_H}=\left(\frac{2(d+z-2)}{d+z-3}\right)^{\frac{1}{d+z-1}},\hspace{2cm}\frac{\rho_c^*}{\rho_H}=2\sqrt{\frac{d+z-1}
{d+z-3}}\left(\frac{d+z-3}
{2(d+z-2)}\right)^{\frac{d+z-2}{d+z-1}}.\ee

Since we are dealing with large entangling regions one can say
that $\rho_{im}=\rho_m $, then after making use of the above
relations one can show that \be\label{ve} v_E=\sqrt{\frac{d+z-1}
{d+z-3}}\left(\frac{d+z-3}
{2(d+z-2)}\right)^{\frac{d+z-2}{d+z-1}}\,\rho_H^{1-z}.\ee  Note
that the mutual information \eqref{muTu} is positive as long as
$\rho_H\ll \frac{h}{2}$ and  $\rho^{z-1}\frac{h}{2}\ll t$,
moreover in this time interval the mutual information for $d+z>3$
is always bigger than $I_{\rm vac}$. It is clear that the mutual
information grows linearly with time. It is also worth mentioning
that the existence of subleading term
$\frac{c_2}{\rho_H^{d-2}}$ in obtaining the entanglement entropy
in this time interval play an important role in getting a positive
mutual information.

In this time interval during the thermalization of system the
mutual information grows linearly with time and there is an upper
limit of the mutual information which takes place when the
entanglement entropy associated with the entangling region $\ell$
saturates to its equilibrium value at $
t\sim\frac{\ell}{2}\rho_H^{z-1}-c_2\rho^z_H+c_{d-1}\frac{\rho_H^{d+z-2}}{\ell^{d-2}}.$
 More precisely setting \be\label{TMAX} v_E \,\,t_{\rm
max}^{(1)}\sim \frac{\ell}{2}-c_2
\rho_H+c_{d-1}\frac{\rho_H^{d-1}}{\ell^{d-2}}, \ee one finds
\be\label{IMax1} I_{\rm max}^{(1)}\approx I_{\rm
vac}+\frac{L^{D-1}}{4G_N}\left(\frac{c_{d-1}}{\ell^{d-2}}-\frac{c_{d-1}}{h^{d-2}}\right)+
\frac{L^{D-1}}{4G_N\rho_H^{d-1}}
\left(\frac{\ell}{2}-\frac{h}{2}\right). \ee Here  $t_{\rm
max}^{(1)}$ is the time when the mutual information reaches its
maximum value $I_{\rm max}^{(1)}$. Note that this maximum value is
independent of $z$.


\subsubsection{ Linear decreasing: $\rho_H^{z-1}\frac{\ell}{2} < t < \rho_H^{z-1}(\ell+\frac{h}{2})$}

In this time interval, one
deduces that both entanglement entropies $S(\ell)$ and $S(h)$
should be approximated by their saturated values (noting that the saturation of $S(\ell)$ takes place at $\frac{\ell}{2}-c_2
\rho_H+c_{d-1}\frac{\rho_H^{d-1}}{\ell^{d-2}}$ and in this time $S(h)$ is already saturated). On the other hand
corresponding entanglement entropy of $2\ell+h$ still grows
linearly with time which is given by \eqref{SBHL}. Plugging these
three entanglement entropies in \eqref{MI} one finds
\be\label{SD0} I=
\frac{L^{D-1}}{4G_N}\left(\frac{c_{d-1}}{(2\ell+h)^{d-2}}-\frac{c_2}{\rho_H^{d-2}}+\frac{\ell}{\rho_H^{d-1}}
-\frac{h}{2\rho_H^{d-1}}-\frac{\sqrt{-f(\rho_{3m})}}{\rho_{3m}^{d+z-2}}t\right),
\ee  which may be simplified as follows \be\label{SD} I\approx
I_{\rm
max}^{(1)}+\frac{L^{D-1}}{4G_N}\left(\frac{c_{d-1}}{\ell^{d-2}}-\frac{c_2}{\rho_H^{d-2}}\right)
+\frac{L^{D-1}}{4G_N\rho_H^{d-1}}\left(\frac{\ell}{2}-v_E
t\right). \ee The mutual information in this time interval
declines linearly  with time and is positive for
$t<\rho_H^{z-1}(\ell+\frac{h}{2})$ and also $I<I_{\rm max}^{(1)}$.


\subsubsection{Saturation}

Final step in thermalization takes place if time passes enough
when all the entanglement entropies $S(\ell_i)$ saturate to their
equilibrium value. In fact the last one is $S(2\ell+h)$ with the
saturation time as given by \be t\sim
(\ell+\frac{h}{2})\rho_H^{z-1}-c_2
\rho_H^z+c_{d-1}\frac{\rho_H^{d+z-2}}{(2\ell+h)^{d-2}}.\ee
Accordingly the mutual information will also saturates to its
equilibrium value. However in Ref.\cite{Alishahiha:2014jxa} it was
shown that the time when the mutual information reaches to its
equilibrium value is not just the same as the saturation time of
$S(2\ell+h)$. But thanks to the proposed conditions in which both
the width of strips $\ell$ and distance between them $h$ are large
compared to the radius of the horizon namely, $\rho_H\ll \ell$ and
$\rho_H\ll h$, the mutual information becomes zero at the end of
the thermalization process. Indeed assuming the mutual information
decreases all the way till it becomes zero, from equation
\eqref{SD0}, one should set \be\label{Isat1}
I^{(1)}_{\text{sat}}\approx\frac{L^{D-1}}{4G_N}\left(\frac{c_{d-1}}{(2\ell+h)^{d-2}}-\frac{c_2}{\rho_H^{d-2}}+\frac{\ell}{\rho_H^{d-1}}
-\frac{h}{2\rho_H^{d-1}}-\frac{v_E}{\rho_H^{d-1}}t^{(1)}_s\right)=0,
\ee so that the saturation time reads \bea\label{tsat}
v_E\,t^{(1)}_s\approx
\ell-\frac{h}{2}-c_2\rho_H+\frac{c_{d-1}\rho_H^{d-1}}{(2\ell+h)^{d-2}},\eea
or for large entangling regions one write  $t^{(1)}_s\approx
(\ell-\frac{h}{2})\rho_H^{z-1}-c_2\rho_H^z$, comparing with the
saturation time of the $S(2\ell+h)$ reveals that the mutual
information saturates earlier than the saturation time of the
entanglement entropy of a strip with width $2\ell+h$.\footnote{ In
Vaidya geometry, similar result has been observed numerically
 in
 \cite{{Balasubramanian:2011at},{Allais:2011ys},{Callan:2012ip},{Li:2013sia}}.}

To summarize the results of first regime in which $t\ll
\rho_H^z$, one sees that the mutual information in the
hyperscaling violating backgrounds after a quantum quench
undergoes four main situations: it starts from its value in the
vacuum and remains almost constant up to $t\sim
\rho_H^{z-1}\frac{h}{2}$, after that it grows with time linearly
till its maximum value which takes place at
$t_{\text{max}}^{(1)}$. Then the mutual information decreases
linearly with time till it becomes zero at the saturation time
which takes place approximately at $t_s^{(1)}\sim
(\ell-\frac{h}{2})\rho_H^{z-1}-c_2\rho_H^z$.


\subsection{Second regime: $\frac{h}{2}\ll\rho_H\ll
\frac{\ell}{2}<\ell+\frac{h}{2}$}

In this regime, to study the behavior of the mutual information,
similar to the previous regime, five time intervals can be
considered separately, as stated below \bea t \ll
\rho_H^{z-1}\frac{h}{2},\hspace{2cm}\frac{h}{2}\rho_H^{z-1} \ll
t\ll\rho_H^z,\hspace{2cm}\rho_H^z \ll t \ll
\rho_H^{z-1}\frac{\ell}{2},\nonumber\\ \rho_H^{z-1}\frac{\ell}{2}<
t< \rho_H^{z-1}(\ell+\frac{h}{2}),\hspace{3.3cm}\mbox{Saturation
time}.\hspace{1.6cm}\eea At the very early time $t \ll
\rho_H^{z-1}\frac{h}{2}$, the behavior of $S(\ell_i)$ is similar
to what discussed in previous case, so that the behavior of the
entanglement entropy at the very early time is given by
\eqref{early}, which means that it remains constant till
$t\ll\rho_H^{z-1} \frac{h}{2}$.

\subsubsection{Non-linear growth:  $\frac{h}{2}\rho_H^{z-1} \ll t\ll\rho_H^z$}

Having noted that $\frac{h}{2}\ll\rho_H$, the corresponding
co-dimension two hypersurface of region $h$ cannot probe the $v<0$
region, namely the corresponding hypersurface remains always at
$v>0$ region which is, indeed, a static hyperscaling violating AdS
black brane. Thus for $S(h)$ one should use its equilibrium value
which is given by equation \eqref{SATH}. However, due to the $t\ll
\rho_H^{z-1}\frac{\ell}{2}$ situation, one can say that $S(\ell)$
and $S(2\ell+h)$ are still at the early times which has been given
by equation \eqref{leading}. Plugging these three entropies in
\eqref{MI} one gets
 \bea\label{quadraticgrowth} {I}\approx I_{\rm vac}
+\frac{L^{D-1}}{4G_N\rho_H^{d-1+z}}\left(-c^z_1h^{z+1}+\frac{(zt)^{1+1/z}}{2z+2}\right),
\eea showing that the mutual information has a non-linear growth
up to $t\sim \rho_H^z$.\\ It is worth mentioning that even though we are dealing with hyperscaling violating geometry, among three free parameters of the theory, namely $D,\,\,\theta$ and $z$, non-linear evolution of mutual information in this time interval is independent of  hyperscaling violating parameter $\theta$. In fact, $\theta$ 
leads to an effective dimension as $D-\theta$ and nontrivial behaviors of the resultant mutual information are the same as that of Lifshitz ($\theta=0,\,\,z\ne0$) geometry in $D+1$ dimensions. On the other hand it was argued that time scaling behavior of the entanglement entropy is dimension-independent \cite{Liu:2013iza}, in this sense one can say that the non-linear behavior of mutual information (scaling as $t^{1+1/z}$) is independent of $\theta$ and one can say $z$ appears in thermodynamical quantities like those in Lifshitz backgrounds.


\subsubsection{Linear growth:  $\rho_H^z \ll t \ll \rho_H^{z-1}\frac{\ell}{2}$}

For $S(h)$, there is no change and hence it is still given by
equation \eqref{SATH}. On the other hand the system has reached a
local equilibrium and due to $\rho_H \ll \frac{\ell}{2}$, the
entanglement entropies of $S(\ell)$ and $S(2\ell+h)$ should be
approximated by equation \eqref{SBHL}. Thanks to the fact that the
entangling regions are large so that $\rho_{i\,m}=\rho_{m}$, one
obtains the following expression for the mutual information \bea
{I}\approx I_{\rm vac}+\frac{L^{D-1}}{4G_N\rho_H^{d-1}} \left( v_E
t-\frac{c^z_1}{4(d-2)\rho^z_H} h^{z+1}\right). \eea In this time
interval, the conditions $h\ll \rho_H$ and $\rho^z_H\ll t$
guarantee that the resultant mutual information will be positive
and bigger than $I_{\rm vac}$. The mutual information linearly
grows with time till when the $S(\ell)$ saturates to its
equilibrium value. Actually this happens at \bea\label{tMax2}
t^{(2)}_{\rm max}\approx \frac{\ell}{2}\rho_H^{z-1}-c_2
\rho^z_H+c_{d-1}\frac{\rho_H^{d+z-2}}{\ell^{d-2}}. \eea  One can
use equation \eqref{tMax2} to estimate the maximum value of the
mutual information as follows \bea\label{IMax2} {I}^{(2)}_{\rm
max}\approx I_{\rm vac}+\frac{L^{D-1}}{4G_N\rho_H^{d-1}} \left(
\frac{\ell}{2}-c_2
\rho_H+c_{d-1}\frac{\rho_H^{d-1}}{\ell^{d-2}}-\frac{c^z_1}{4(d-2)\rho^z_H}h^{z+1}
\right). \eea


\subsubsection{Linear decreasing: $\rho_H^{z-1}\frac{\ell}{2}< t< \rho_H^{z-1}(\ell+\frac{h}{2})$}

The main change in the present time interval comes from the fact
that entanglement entropy $S(\ell)$ is saturated to its
equilibrium value. However, due to the size of entangling region
$\ell$, one should follow the situation which has been already
discussed in appendix to estimate the corresponding entanglement
entropy which is given by \be S(\ell)\approx\frac{L^{D-1}}{4G_N
}\left(\frac{1}{(d-2)\epsilon^{d-2}}+\frac{\ell}{2\rho_H^{d-1}}
-\;\frac{c_2}{\rho_H^{d-2}}\right). \ee Noting that there is no
change in reading the $S(2\ell+h)$ and $S(h)$ which are given by
\eqref{SBHL} and \eqref{SATH} respectively. Therefore, the mutual
information has a decreasing regime in this time interval as
stated below \bea\label{YY} {I}\approx I^{(2)}_{\rm max}
+\frac{L^{D-1}}{4G_N}\left(\frac{c_{d-1}}{\ell^{d-2}}-\frac{c_2}{\rho_H^{d-2}}\right)+\frac{L^{D-1}}{4G_N\rho_H^{d-1}}\left(\frac{\ell}{2}-v_E
t\right). \eea Note that the mutual information is positive and
also $I<I^{(2)}_{\rm max}$.


\subsubsection{Saturation}

After a long time the system which we are dealing with reaches to
thermal state whose gravity dual is provided by a hyperscaling
violating AdS black brane. However, as previous case having noted
the condition $\frac{h}{2}\ll \rho_H\ll \frac{\ell}{2}$, one can
use \eqref{MIS} for the mutual information and hence the
equilibrium value of the mutual information becomes
\bea\label{satcase2} I^{(2)}_{\text{sat}}\approx
\frac{L^{D-1}}{4G_N}\left(\frac{c_{d-1}}{h^{d-2}}-\frac{c_2}{\rho_{H}^{d-2}}-\frac{h}{2\rho_H^{d-1}}-\frac{c^z_1}{4(d-2)\rho_H^{d-1+z}}h^{z+1}\right).
\eea The above expression for mutual information can be written as
\be\label{Isat2}
I^{(2)}_{\text{sat}}=I_{\text{vac}}-\frac{L^{D-1}}{4G_N}\left(\frac{c_2}{\rho_{H}^{d-2}}+\frac{h}{2\rho_{H}^{d-1}}+
\frac{c^z_1}{4(d-2)\rho_H^{d-1+z}}h^{z+1}+\frac{c_{d-1}}{(2\ell+h)^{d-2}}-\frac{2c_{d-1}}{\ell^{d-2}}\right)
\ee which shows that in this case with the condition
$\frac{h}{2}\ll\rho_H\ll \frac{\ell}{2}$ the expression in the
parentheses is always positive leading to the fact that
$I^{(2)}_{\text{sat}}<I_{\text{vac}}$.

In order to estimate the time in which the mutual information
saturates in this regime, we should just follow the recipe in
which the mutual information decreases linearly with time till it
reaches its equilibrium value. In other words the saturating time
can be obtained by equating equations \eqref{YY} and
\eqref{satcase2}, and making use of \eqref{ve}, one finds the
saturation time as \be t^{(2)}_s\approx
(\ell+\frac{h}{2})\rho_H^{z-1}-c_2\rho_H^z.\ee

Let us summarize the results of present regime
$\frac{h}{2}\ll\rho_H\ll \frac{\ell}{2}<\ell+\frac{h}{2}$. The
time evolution of the mutual information undergoes five main
phases: it starts from the value in vacuum and remains almost
constant up to $t\sim \rho_H^{z-1}\frac{h}{2}$, then it grows
non-linearly with time as $t^{1+1/z}$ till $t\sim \rho^z_H$. After
that it linearly grows till its maximum value which takes place at
$t_{\text{max}}^{(2)}$. After the maximum value, it decreases
linearly with time and finally it saturates to a constant value at
the saturation time $t^{(2)}_s$.


\subsection{Third regime: $\frac{h}{2}\ll \frac{\ell}{2}<\rho_H<\ell+\frac{h}{2}$}

In this regime the related co-dimension two hypersurfaces of
$S(h)$ and $S(\ell)$ cannot probe the region near and behind the
horizon and entangling regions saturate to their equilibrium
values before the system reaches a local equilibrium. Thus only
$S(2\ell+h)$ grows linearly with time before it reaches its
equilibrium value.\\
The very early time behavior is in fact the same as the similar
case of the previous regime, so that one can say that the mutual
information has a fixed value in the vacuum till $t\sim
\rho_H^{z-1}\frac{h}{2}$ and then it begins to grow non-linearly
with time as stated below

\bea \label{nonline}{I}\approx I_{\rm vac}
+\frac{L^{D-1}}{4G_N\rho_H^{d-1+z}}\left(-c^z_1h^{z+1}+\frac{(zt)^{1+1/z}}{2z+2}\right).
\eea

To find the maximum value of the mutual information in this
regime, the time behavior of $S(\ell)$ plays a key role, it
actually grows non-linearly till reaches to its equilibrium value
which is given by \eqref{SATH}, therefore one obtains an
estimation for the time when the mutual information becomes
maximum as follows \be t^{(3)}_{\text{max}}\sim
\frac{1}{z}\left(\frac{2z+2}{4(d-2)}c_1^z\right)^{\frac{z}{z+1}}\ell^z.\ee
By making use of this maximum time one can obtain the maximum
value of the mutual information as \be\label{IMax3} I^{(3)}_{\rm
max}\approx I_{\rm vac}
+\frac{L^{D-1}c^z_1}{4G_N\rho_H^{d-1+z}}\left(\frac{\ell^{z+1}}{4(d-2)}-h^{z+1}\right).\ee
Let us now study the other  time intervals in more details.


\subsubsection{Non-linear decreasing:  $\frac{\ell}{2}\rho^{z-1} < t < \rho^z_H$}

As already mentioned in this time interval the entanglement
entropies of $S(h)$ and $S(\ell)$ are saturated to their
equilibrium values which are given by \eqref{SATH}, whereas
$S(2\ell+h)$ is still at the early times and should be
approximated by equation \eqref{leading}

\be S(2\ell+h)\approx
\frac{L^{D-1}}{4G_N}\left[\frac{1}{(d-2)\epsilon^{d-2}}-\frac{c_{d-1}}{(2\ell+h)^{d-2}}+\frac{m}{2(z+1)}(zt)^{1+\frac{1}{z}}\right].\ee
Plugging these results into equation \eqref{MI}, one finds \bea
{I}\approx I_{\rm max}^{(3)}+\frac{L^{D-1}}{4G_N\rho_H^{d-1+z}}
\left(\frac{ c^z_1}{2(d-2)}\ell^{z+1}-
\frac{(zt)^{1+1/z}}{2(z+1)}\right). \eea Note that since
$t>t^{(3)}_{\rm max}$ it is clear that  $I<I^{(3)}_{\rm max}$.


\subsubsection{Linear decreasing:  $ \rho^z_H< t<\rho_H^{z-1}(\ell+\frac{h}{2})$}

There is no change for the entanglement entropies  $S(h)$ and
$S(\ell)$, however, $S(2\ell+h)$ is locally equilibrated as
already discussed in \eqref{SBHL}. Making use of the large
entangling region limit for $\rho_{m}$, one can show that the
mutual information reads as \bea\label{SD3} {I}\approx
I^{(3)}_{\rm max}+\frac{L^{D-1}}{4G_N} \left(
\frac{c^z_1}{4(d-2)\rho_H^{d-1+z}} \ell^{z+1}-v_E t\right). \eea
We note that in this time interval the mutual information is
positive as long as $t <\rho_H^{z-1}( \ell+\frac{h}{2})$.


\subsubsection{Saturation}

Again if one wait enough the system is going to be saturated. But
in this time interval the entanglement entropies $S(h)$ and
$S(\ell)$ have already been saturated here is no change in reading
them and given by equation \eqref{SATH}, though the entanglement
entropy $S(2\ell+h)$ must be approximated by equation
\eqref{EEBHh}. Therefore the mutual information can be recast as
\be\label{satcase3} I_{\text{sat}}^{(3)}\approx
I_{\text{vac}}+\frac{L^{D-1}}{4G_N}\left(\Big(\frac{c_1^z}{4(d-2)\rho_H^{d-1+z}}\Big)(2\ell^{z+1}-h^{z+1})-\frac{2\ell+h}{2\rho_H^{d-1}}+\frac{c_2}{\rho_H^{d-2}}
-\frac{c_{d-2}}{(2\ell+h)^{d-2}}\right). \ee In the present regime, one can show that $I_{\text{sat}}^{(3)}<I_{\text{vac}}$. In order to estimate the saturation time, with the assumption of large
entangling region and noting that the mutual information decreases
linearly with time, one may equate equations \eqref{satcase3} and
\eqref{SD3} to find \be\label{tsat3}
t_s^{(3)}\sim(\ell+\frac{h}{2})\rho_H^{z-1}-c_2\rho_H^z.\ee

To summarize the results of the third regime $\frac{h}{2}\ll
\frac{\ell}{2}<\rho_H<\ell+\frac{h}{2}$, what we have obtained for
the time evolution of the mutual information is as follows: at the
very early time till $t\sim\rho_H^{z-1} \frac{h}{2}$, it has a
constant value as it was in the vacuum state, then from
$t\sim\rho_H^{z-1} \frac{h}{2}$ to $t_{\text{max}}^{(3)}$ it has
non-linear growing behavior. After its maximum we have two
declining phases non-linear which is followed by linear phase just
before saturation. The mutual information saturates to a constant
value at the saturation time $t_s^{(3)}$.


\subsection{Fourth regime: $\frac{h}{2}\ll \frac{\ell}{2}<\ell+\frac{h}{2}\ll \rho_H$}

The final case takes place when we are interested in a condition
that all the entanglement entropies saturate to their equilibrium
values before the system reaches a local equilibrium. The
situation is very similar to the third regime, namely mutual
information starts from its value in vacuum and remains fixed up
to $t\sim \rho_H^{z-1}\frac{h}{2}$ that it starts growing
non-linearly with time which could be estimated as relation
\eqref{nonline}.  This non-linear behavior lasts up to its
maximum value which is given by \be t^{(4)}_{\text{max}}\sim
\frac{1}{z}\left(\frac{2z+2}{4(d-2)}c_1^z\right)^{\frac{z}{z+1}}\ell^z.\ee
Similarly the maximum value reads as \be\label{I4max} I^{(4)}_{\rm
max}\approx I_{\rm vac}
+\frac{L^{D-1}c^z_1}{4G_N\rho_H^{d-1+z}}\left(\frac{\ell^{z+1}}{4(d-2)}-h^{z+1}\right).\ee
After that the thermalization is followed by a nonlinear
decreasing as \bea \label{I4}{I}\approx I_{\rm
max}^{(4)}+\frac{L^{D-1}}{4G_N\rho_H^{d-1+z}} \left(\frac{
c^z_1}{2(d-2)}\ell^{z+1}- \frac{(zt)^{1+1/z}}{2(z+1)}\right). \eea
Noting that since $t>t^{(4)}_{\rm max}$ again one gets
$I<I^{(4)}_{\rm max}$.

Finally after a long time the mutual information reaches its
equilibrium value and the saturation takes place when all the
entanglement entropies become that of a hyperscaling violating AdS
black brane which is given by equation \eqref{SATH}. Therefore
from equation \eqref{MI}, the saturated mutual information is
obtained as \be\label{I4sat}
 I_{\rm sat}^{(4)}\approx I_{\rm vac}-\frac{L^{D-1}}{16G_N(d-2)}
 c^z_1\frac{(\ell+h)^{z+1}}{\rho_H^{d-1+z}}.
\ee The saturation time can be estimated easily by assuming that
the non-linear decreasing continues all the way to the
saturation point then after equating Eq.s \eqref{I4} and
\eqref{I4sat} one can estimate the saturation time.

\section{Conclusion}

In this paper we studied the thermalization of the mutual
information after a global quantum quench in a hyperscaling
violating background. Within the language of the gauge/gravity
duality a global quantum quench for a strongly coupled field
theory with hyperscaling violation and an anisotropic scaling
symmetry may be described by an AdS-Vaidya geometry with a
hyperscaling violating factor at the gravity side and the quantum
quench might be thought as an instant injection of matter in a
small time interval. We used the covariant prescription for
computing the holographic entanglement entropy and studied mutual
information, holographically. This was done by extremizing a
certain codimension-two hypersurface in the bulk whose metric is
given by hyperscaling violating AdS- Vaidya geometry. In this
study following Ref.\cite{Alishahiha:2014jxa} we used two parallel
strips with width $\ell$ separated by distance $h$ assuming that
$h\ll\ell$. Actually thermalization depends on the evolution of
the corresponding hypersurfaces of the entanglement entropies
which is controlled by the region inside and around the horizon.
Beside the radius of horizon the size of entangling
regions can be used as a scale in probing the system. In other
words the time behavior of the thermalization depends on relative
size of the corresponding entangling regions and the radius of
horizon. In fact depending that the size of entangling regions
could be larger or smaller than the radius of horizon one may
distinguish four main regimes. It goes as a general rule in which
if the width of entangling region is smaller than the radius
of horizon, the corresponding entanglement entropy grows
non-linearly with time as $t^{1+1/z}$ and saturates before the
system reaches a local equilibrium. Noting that locally
equilibrium occurs when the system has ceased production of
thermodynamic entropy and the entanglement entropy can be given in
terms of the thermal entropy. Another important case takes place
when the width of the entangling region is larger than the radius
of horizon, the corresponding entanglement entropy grows
non-linearly with time before the system reaches a local
equilibrium, with a linear growth after local equilibrium, it
saturates. \\  It was argued that the early time behavior of the
entanglement entropy depends on $z$ while in the intermediate
region its time behavior is linear \cite{Alishahiha:2014cwa}. We
extended these statements for the thermalization of mutual
information. We found that except the change in the time scale due
to the different scaling of time, the very early time behavior is
independent of $z$ and is similar to what done in
Ref.\cite{Alishahiha:2014jxa}, however, the non-linear behavior is
contorted by $z$ and with $z=1$ one covers the previous results of
quadratic behavior. On the other hand for large $z$ the non-linear
behavior turns to a linear behavior.

Actually understanding how quantum information spread in a strongly coupled system which is out of equilibrium is a question of much importance in many different areas of physics. In a 2-dimensional CFT, Calabrese and Cardy presented a simple physical picture for the linear growth behavior and saturation of the entanglement entropy \cite{CC}. In this model, entanglement entropy spreads via free propagation of EPR pairs of entangled \emph{quasiparticles}, these 'particles' is assumed to be created by the injected energy density due to a global quench
at $t = 0$ and subsequently propagate freely with the speed of light. The gravity dual of this picture has been analyzed in \cite{Hartman:2013qma}. Moreover, for a strongly coupled system with a gravity dual, the so-called entanglement \emph{tsunami} proposal introduced in \cite{Liu:2013iza, Liu:2013qca, Leichenauer:2015xra} could address  the linear and quadratic behaviors of the entanglement entropy after a global quantum quench \cite{Casini:2015zua}. In this model the growth in entanglement entropy is described via a sharp wave-front carrying entanglement inward from the boundary of entangling region. Although the time scaling behavior of the entanglement entropy after a quench seems to be an open question, in the hyperscaling violating backgrounds, the linear growth behavior of the entanglement entropy is recovered, however, scaling like $t^{1+\frac{1}{z}}$ indicates that at early time after quench the entanglement entropy is sensitive to the initial state.

In the literature of quantum information the subjects say as
$n$-partite information or multi-partite entanglement become
important \cite{Horodecki:2009zz}. When we are interested in
measurement of the amount of information or correlations (both
classical and quantum) between $n$ disjoint regions
$A_i,\;i=1,\cdots, n$, the $n$-partite information may provide a
good model for studying this process. It is shown that both
$n$-partite information or multi-partite entanglement can be
written in terms of the mutual information. The results of present
work can be extended to those subjects to study their
non-equilibrium behaviors.


\section{Acknowledgments}

 I would like, first of all, to thank M. Alishahiha, I learned
 a lot from his weekly meetings. I am really indebted to him. Without discussions with M. Reza
 Mohammadi Mozaffar this paper would probably not exist, hereby I also thank him.  I would also like to acknowledge and thank  A. Mollabashi, A. Faraji, F.
Omidi and A. Naseh for some related discussions. This work has
been supported in parts by Islamic Azad University Central Tehran
Branch.


\appendix
\section{Appendix} \label{App:AppendixA}

As it is known in a given system with Hilbert space $H$, the
states are characterized by the density matrix $\rho$. Now suppose
this system can be divided by two subsystems $A$ (which mostly
supposed as a spatial region) and $\bar{A}$ which stands for the
complement of $A$, so that one can write $H=H_A \otimes
H_{\bar{A}}$. The reduced density matrix of $A$ can be obtained by
tracing over the degrees of freedom of $\bar{A}$, then the
entanglement entropy of $A$ is defined by the von Neumann entropy
of the corresponding reduced density matrix: $S_A=-{\rm Tr}(\rho_A
\log\rho_A)$. Entanglement entropy has a UV divergent term where
in spatial dimensions bigger than one, the divergent term obeys
the area law, namely it is proportional to the area of the
entangling region \cite{Srednicki:1993im}, while for two
dimensional spacetimes the divergent term is logarithmic (see for
example \cite{{Holzhey:1994we},{Calabrese:2004eu}} for two
dimensional CFT). In the context of the quantum field theory the
replica trick has been mostly used to obtain the entanglement entropy which is based on computing the Renyi entropies (for details see
\cite{Calabrese:2009qy}). However, for theories which have gravity
dual, the Ryu-Takayanagi conjecture has been used to calculate the
entanglement entropy. Noting that in the case of the time-dependent backgrounds one
should use the covariant proposal of the entanglement
entropy\cite{Hubeny:2007xt}.

\subsection{Static Hyperscaling Violating Backgrounds}

According to the holographic description for a given entangling
region, the entanglement entropy is given by \be S=\frac{{\cal
A}}{4G^{D+2}_N},\ee where $G_N$ is the Newton constant and ${\cal
A}$ is the $D-$dimensional minimal surface in the bulk whose
boundary coincides with the boundary of the entangling region. Let
us choose a strip at a fixed time on the boundary as an entangling
region as indicated below \be \label{region}-\frac{\ell}{2}\leq
x_1\equiv x\leq \frac{\ell}{2},\hspace{1cm}0\leq x_a\leq L,\,\,\,
\,\,\,\,a=2,\cdots,D,\ee where $(t, \vec{x})$ are the space-time
coordinates. Now the aim is to find the surface in the bulk with
the boundary of the above strip and then minimize it. The
corresponding metric is given by \eqref{met1} and the profile of
the hypersurface in the bulk may be parameterized by $x(r)$, after
setting $r=\rho^{-1},$ the induced metric on this hypersurface is
given by \be
ds_{\mbox{ind}}^2=\rho^{2\frac{\theta}{D}-2}\bigg[\Big(\frac{1}{f(\rho)}+x'^2\Big)d\rho^2+\sum_{a=2}^D dx_a^2\bigg]\ee
where here the prime stands for the derivative with respect to
$\rho$ and $f(\rho)$ is as follows \be \label{f(rho)}
f(\rho)=1-m\rho^{d-1+z}=1-(\frac{\rho}{\rho_H})^{d-1+z}.\ee Thus
the area reads as \be\label{area} {\cal A}=\frac{L^{D-1}}{2}\int
d\rho\frac{\sqrt{f^ {-1}+x'^2}}{\rho^{d-1}},\ee where one should
minimize the area and hence one obtains \be\label{l/2}
\frac{\ell}{2}=\int_0^{\rho_t}d\rho\frac{(\frac{\rho}{\rho_t})^{d-1}}{\sqrt{f(\rho)\Big(1-(\frac{\rho}{\rho_t})^{2(d-1)}\Big)}},\ee
note that $\rho_t$ is the turning point of the extremal
hypersurface in the bulk. Therefore, if one introduces $\epsilon$
to be a UV cutoff of theory the entropy becomes \be\label{s}
S=\frac{L^{D-1}}{4G_N}\int_\epsilon^{\rho_t}d\rho\frac{1}{\rho^{d-1}\sqrt{f(\rho)\Big(1-(\frac{\rho}{\rho_t})^{2(d-1)}\Big)}}.\ee
In general there is no explicit analytic expression for the
entanglement entropy, however, in some certain limits say as
$m\ell^{d-1+z}\ll 1$, one can expand the area \eqref{area} around the
$f=1$ which leads to the following expression for the change of
the entanglement entropy \begin{equation}
 \Delta S=S_{BH}-S_{vac}=\frac{L^{D-1}}{16G_N(d-2)} c^z_1\frac{\ell^{z+1}}{\rho_H^{d-1+z}},\label{SATH}
\end{equation} where \be
\label{c1z}c_1=\frac{\Gamma(\frac{1}{2(d-2)})}{2\sqrt{\pi}\Gamma(\frac{d}{2(d-1)})}.\ee
change of the geometry between black hole solution $f\neq 1$
and vacuum solution $f=1$. For a strip the entanglement
entropy for vacuum was found in \cite{Dong:2012se} \be\label{SV}
S_{\text{vac}}=\left\{\ba{ll}
\frac{L^{D-1}}{4G_N}\left(\frac{1}{(d-2)\epsilon^{d-2}}-\frac{c_{d-1}}{\ell^{d-2}}\right)
&{\rm for}\;\;d> 2,\cr &\cr \frac{L^{D-1}}{4G_N}\ln
\frac{\ell}{\epsilon},&  {\rm for}\;\;d= 2, \ea \right. \ee where
\be
\label{cdef}c_{d-1}=\frac{2^{d-2}}{d-2}\left(\frac{\sqrt{\pi}\Gamma(\frac{d}{2(d-1)})}{\Gamma(\frac{1}{2(d-1)})}\right)^{d-1}.\ee

On the other hand for large entangling regions $m\ell^{d-1+z}\gg 1$, the
main contribution comes from the limit where the minimal surface
is extended to the horizon, namely $\rho_t\sim\rho_H$. In this
limit after making use of the expansion of the equations
\eqref{l/2} and \eqref{s} one finds \be
S_{BH}\sim\frac{L^{D-1}}{4G_N}\frac{1}{\rho_H^{d-1}}\,\frac{\ell}{2}+\frac{L^{D-1}}{4G_N}\frac{1}{\rho_H^{d-2}}\int_{\frac{\epsilon}{\rho_H}}^1
\frac{\sqrt{1-\xi^{2(d-1)}}\,d\xi}{\sqrt{1-\xi^{d-1+z}}\,\xi^{d-1}},\ee
where $\xi$ is a dimensionless parameter defined by
$\xi\equiv\frac{\rho}{\rho_H}$.
The above integral can not be solved,
however by extracting its divergence, it can be written as
\be\label{regu}
\int_{\frac{\epsilon}{\rho_H}}^1\,\,\frac{d\xi}{\xi^{d-1}\sqrt{(1-\xi^{d-1+z})(1-\xi^{2(d-1)})}}=\frac{1}{(d-2)\epsilon^{d-2}}-c_2,\ee
where for a fixed $z$ and $d$ it can be computed, as shown in
below
\begin{table}[h]
\caption{$c_2$ in $S_{BH}$ with certain $z$ and $d$} \vspace{5pt}
  \centering
 \begin{tabular}{c||c|c|c|c|c}
\hline
\head{\shortstack{\\ $z$ and $d$}} & \head{$z=1,\,\,d=3$} & \head{$z=1,\,\,d=4$} & \head{$z=1,\,\,d=5$} & \head{$z=2,\,\,d=3$}& \head{$z=2,\,\,d=4$}\\[5pt]
\hline\\
$c_2$ & 0.88 & 0.33 & 0.14 & 1 & 0.44\\[5pt]
\hline
\end{tabular}\label{table}
\end{table}

Therefore the entanglement entropy reads as \be\label{EEBHh}
S_{BH}\approx\frac{L^{D-1}}{4G_N}\left(\frac{1}{(d-2)\epsilon^{d-2}}+\frac{\ell}{2\rho_H^{d-1}}-\frac{c_2}{\rho_H^{d-2}}\right).\ee

\subsection{Time-dependent Hyperscaling Violating Backgrounds}

Entanglement entropy could potentially provide useful quantity of
time scaling in a system undergoing a rapid change (quantum
quench) which this change brings the system out of equilibrium. In
our case in hand, as mentioned previously, at the gravity side the
Vaidya metric can well describe an infalling shell of a massless
and pressureless matter in a hyperscaling violating geometry. The
corresponding metric is \be \label{met2}
ds_{D+2}^2=\rho^{2\frac{\theta-D}{D}}\Bigg(-\rho^{2-2z}f(\rho,v)dv^2-2\rho^{1-z}d\rho
dv+\sum_{i=1}^Ddx_i^2\Bigg), \ee in which we have employed the
Eddington-Finkelstein- like coordinate in \eqref{met1} as defined
by \be dv=dt+\frac{dr}{f(r)r^{z+1}},\ee which followed by
setting $r=\rho^{-1}$ (for mathematical details see
\cite{Alishahiha:2014cwa} and Ref.s therein). With the effective
dimension one has $f(\rho,v)=1-m(v)\rho^{d-1+z}.$

For a system after a global quantum quench, the entanglement
entropy could provide a proper scaling during the process of the
thermalization. In the process of evolution two phases become
crucial: a time growing phase and a saturation phase where the
entanglement entropy saturates to its equilibrium value. The
radius of the horizon $\rho_H$, and the size of the entangling
region can be used as scales to address these two phases. For a
strip as an entangling region, the size is given by its width
$\ell$. We use a strip with width $\ell$ defined in
\eqref{region} as an entangling region. Following the holographic
description, in computing the entanglement entropy one should use
the covariant proposal of finding the extremum area of a
codimension-two hypersurface whose boundary coincides with the
above mentioned strip \cite{Hubeny:2007xt}. At this stage one only
needs to find the extremal surface of the codimension-two
hypersurface in the bulk. To do so, we use the $v(x)$ and
$\rho(x)$ to parameterize the corresponding codimension-two
hypersurface in the bulk. Using the metric \eqref{met2} which is
the bulk metric, one finds the induced metric on the hypersurface
as \be
ds^2_{\mbox{ind}}=\rho^{2\frac{1-d}{D}}\bigg[\Big(1-\rho^{2-2z}f(\rho,v)v'^2-2\rho^{1-z}v'\rho'\Big)dx^2+\sum_{a=2}^D
dx_a^2\bigg], \ee noting that prime represents derivative with
respect to $x$. With this induced metric, the hypersurface area
becomes \be\label{area1} {\cal
A}=\frac{L^{D-1}}{2}\int_{-\ell/2}^{\ell/2}dx\frac{\sqrt{1-2\rho^{1-z}v'\rho'-\rho^{2-2z}v'^2f}}{\rho^{d-1}},
\ee which should be extremized. However, the area expression
\eqref{area1} can be treated as an action of a one-dimensional
quantum mechanical system where the $v(x)$ and $\rho(x)$ are its
dynamical fields. The supposed action is independent of $x$ and
hence the corresponding Hamiltonian becomes a constant of motion
and results in a conservation law \be \label{cont}
H^{-1}\equiv-\rho^{d-1}\sqrt{1-2\rho^{1-z}v'\rho'-\rho^{2-2z}v'^2f}=\mbox{const}.\ee Therefore, the
equations of motion read as \be
\partial_xP_v=\frac{P_\rho^2}{2}\frac{\partial
f}{\partial v},\hspace{1cm}
\partial_xP_\rho=\frac{P_\rho^2}{2}\frac{\partial
f}{\partial\rho}+\frac{d-1}{\rho^{2d-1}}H^{-2}+\frac{1-z}{\rho^{2-z}}P_\rho
P_v,\label{equs}\ee where $P$'s stand for the the momenta
conjugate defined by (up to a factor of $H$)
$$P_\rho=\rho^{1-z}v',\hspace{1cm}P_v=\rho^{1-z}(\rho'+\rho^{1-z}v'f).$$
If $(\rho_t,v_t)$ is the turning point of the extremal
hypersurface in the bulk one can use the following boundary
conditions
\begin{equation}
\rho(0)=\rho_t,\hspace{5mm}v(0)=v_t,\hspace{5mm}v(\frac{\ell}{2})=t,\hspace{1cm}\rho(\frac{\ell}{2})=\rho'(0)=v'(0)=0.
\label{boun}\end{equation}

In general there is no analytic solution for Eq.s \eqref{equs}
but, in a special form of $m$ say as $m(v)=m\,\,\theta(v)$ one
can solve the equations, where $\theta(v)$ is the step function
and therefore for $v<0$ the geometry \eqref{hyper} is an AdS
metric while for $v>0$ it is an AdS-Schwarzschild black hole whose
horizon is located at $\rho_H=m^{-1/(D+1)}$. Quench or a sudden
change in a strongly coupled field theory could be modeled by the
$\theta$-function in the gravity side, let suppose it as follows
\be f(\rho,v)=1-\theta(v)(\frac{\rho}{\rho_H})^{d-1+z}.\ee Now
the aim is to solve Eq.s\eqref{equs}, however because of the step
function, three region can be distinguished as stated below:
\vspace{4mm}

\begin{itemize}
  \item Region I:$\,\,\,\,\,v<0$

In this region theta-function is zero so that $f(\rho, v)=1$
this means that the system is in its vacuum state. The
corresponding gravity dual is given by \be
ds^2=\rho^{\frac{2(1-d)}{D}}\Big(-\rho^{2-2z}dv^2-2\rho^{1-z}d\rho
dv+d\vec{x}^2\Big).\ee Having noted that $\frac{\partial
f(\rho,v)}{\partial v}=0$, from the first equation of
\eqref{equs} and the boundary condition \eqref{boun}, the momentum
conjugate of $v$ is found to be a constant equal to zero \be
P_{(1)v}=\rho^{1-z}(\rho'+\rho^{1-z}v')=0.\ee
Note that we use the index (1) referring the value of quantities in the first case namely $v<0$ region. From the above equation and after making use of the relation
\eqref{cont}, one obtains the following profile of the extremal
surface \be v(\rho)=v_t+\frac{1}{z}(\rho_t^z-\rho^z),
\hspace{15mm} x(\rho)=\int_\rho^{\rho_t}\frac{
\xi^{d-1}\,\,d\xi}{\sqrt{\rho_t^{2(d-1)}-\xi^{2(d-1)}}}. \ee At the null
shell one has $v=0$, consequently from the above equation one
can obtain the point where hypersurface intersects null
shell as \be\rho_c^z=\rho_t^z+zv_t.\ee On the other hand one can
also obtain
\be\rho'_{(1)}=-\rho_c^{1-z}v'_{(1)}=-\sqrt{(\frac{\rho_t}{\rho_c})^{2(d-1)}-1}.\ee\vspace{4mm}
Note that $(\rho_t,v_t)$ refers to the turning point of the
extremal hypersurface in the bulk whereas $(\rho_c,v_c)$ is the
crossing point of the hypersurface and null shell.

\item Region II:$\,\,\,v>0$

In this region one has
$f(\rho)=1-(\frac{\rho}{\rho_H})^{d-1+z}\equiv 1-g(\rho)$ and
hence the corresponding geometry becomes a static black brane
which is given by \be
ds^2=\rho^{\frac{2(1-d)}{D}}\Big(-\rho^{2-2z}f(\rho)dv^2-2\rho^{1-z}d\rho\,dv+d\vec{x}^2\Big).\ee
Once again one has $\frac{\partial f}{\partial v}=0$, so that the corresponding momentum conjugate of $v$ becomes a constant \be
P_{(2)v}=\rho^{1-z}(\rho'+\rho^{1-z}v'f(\rho))=\mbox{const}.
\ee Similarly, from the conservation law \eqref{cont} it can be
shown \be\label{veff}
\rho'^2=\frac{P^2_{(2)v}}{\rho^{2-2z}}+\Big((\frac{\rho_t}{\rho})^{2(d-1)}-1\Big)f(\rho)\equiv
V_{eff}(\rho),\ee where $V_{eff}(\rho)$ might be considered as an
effective potential describing a one-dimensional dynamical system
with a dynamical variable $\rho$, this assumption helps us in
describing the behavior of the entanglement entropy. Using this
effective potential one obtains \be
\frac{dv}{d\rho}=-\frac{1}{\rho^{2(1-z)}f(\rho)}\Big(\rho^{1-z}+\frac{P_{(2)v}}{\sqrt{V_{eff}(\rho)}}\Big).\ee

\item Null Shell $\,\,\,v=0$

At the null shell, one should consider the matching of the results
of two previous regions, noting that $\rho$ and $v$ which are
the space-time coordinates are indeed continues across the null
shell. It is worth to mention that since matter has been injected
along $v$ namely the null direction, the corresponding momentum
conjugate jumps once one moves from the initial phase $(v>0)$ to
the final phase $(v<0)$. However, the momentum conjugate of
$\rho$ remains continuous which means $v'_{(1)}=v'_{(2)}$.
Taking the integration of the equations of motion across the null
shell results in \be
\rho'_{(2)}=\Big(1-\frac{1}{2}g(\rho_c)\Big)\rho'_{(1)},\hspace{5mm}
{\cal L}_{(1)}={\cal L}_{(2)}.\ee Thus the momentum conjugate of
$v$ becomes:\be
P_{(2)v}=\frac{1}{2}\rho_c^{1-z}g(\rho_c)\rho'_{(1)}=-\frac{1}{2}\rho_c^{1-z}g(\rho_c)\sqrt{(\frac{\rho_t}{\rho_c})^{2(d-1)}-1}.\ee

To compute the extremal hypersurface in the bulk, it is important
to mention that the extremal hypersurface could be in fact
extended in both $v<0$ and $v>0$ regions. Accordingly the
width $\ell$ and the boundary time are given by
\be\label{earlytime}
\frac{\ell}{2}=\int_{\rho_c}^{\rho_t}\frac{d\rho\,\,\rho^{d-1}}{\sqrt{\rho_t^{2d-2}-\rho^{2d-2}}}+\int_0^{\rho_c}{\frac{d\rho}{\sqrt{V_{eff}}}},\hspace{15mm}
t=\int_0^{\rho_c}\frac{d\rho\,\,\rho^{z-1}}{f(\rho)}\Big(1+\frac{E\rho^{z-1}}{\sqrt{V_{eff}(\rho)}}\Big),\ee
where $E=P_{(2)v}\rho_t^{z-1}$. Therefore the area of the
hypersurface in the bulk is found to be \be\label{an} {\cal
A}_{d-1}=\frac{L^{D-1}}{\rho_t^{d-2}}\Big(\int_{\frac{\rho_c}{\rho_t}}^1\frac{d\xi}{\xi^{d-1}\sqrt{1-\xi^{2(d-1)}}}+\int_0^{\frac{\rho_c}{\rho_t}}\frac{d\xi}
{\xi^{2(d-1)} \sqrt{R(\xi)}}\Big),\ee where \be R(\xi)\equiv
V_{eff}(\rho_t\xi)=E^2\xi^{2(z-1)}+\Big(\frac{1}{\xi^{2(d-1)}}-1\Big)f(\rho_t\xi),\ee
in which
$f(\rho_t\xi)=1-(\frac{\rho_t}{\rho_H})^{d-1+z}\xi^{d-1+z}$.
Clearly the area \eqref{an} for large volume is divergent (UV
effect) and a proper UV cutoff is actually needed. In fact what we
are interested in is the change of the area when the system
evolves from its vacuum state to an excited state namely,
corresponding area of the extremal hypersurface in the vacuum
solution is given by \be {\cal
A}_{d-1}^{\mbox{vac}}=\frac{L^{D-1}}{\rho_t^{d-2}}\int_0^1\frac{d\xi}{\xi^{d-1}\sqrt{1-\xi^{2d-2}}},\ee
which it could play the role of a regulator. One can use the
expression for $t$,$\,\,\ell$ and ${\cal A}$ to study the scaling
behavior of the entanglement entropy during the process of the
thermalization after a global quantum quench.
\end{itemize}

\subsection{Saturation: late time equilibrium}
After a long time from a change due to the quench, system reaches
to the thermal equilibrium and hence the entanglement entropy
saturates to its equilibrium value. In fact the system is locally
equilibrated for $t\geq \rho_H$ and the saturation takes place
when the extremal hypersurface is entirely outside the horizon
namely $\rho_t<\rho_H$. In the saturation time as long as we are
dealing with the large entangling region, the main contribution of
the hypersurface in the bulk comes from the geometry around the
horizon, namely one has $\rho_c\simeq\rho_t\simeq\rho_H$. Now the
aim is to compute the area and the entanglement entropy in this
limit. By taking $\rho_c=\rho_t(1-\delta)$ for $\delta\ll1$ and
also noting that $P_{(2)v}=0$ and also $\rho_t\simeq\rho_H$, one
can expand \eqref{earlytime} and \eqref{an} in this limit as
\be\frac{\ell}{2}\approx\rho_H\int_0^{1-\delta}\frac{d\xi}{\sqrt{V_{eff}(\rho_t\xi)}},\hspace{2cm}{\cal
A}\approx\frac{L^{D-1}}{\rho_H^{d-2}}\int_{\frac{\epsilon}{\rho_H}}^{1-\delta}\frac{d\xi}{\xi^{2(d-1)}{\sqrt{V_{eff}(\rho_t\xi)}}}.\ee
Noting that the main contribution to the $\ell$ and ${\cal A}$
come from the $\xi=1$, then one can write \be {\cal
A}\approx\frac{L^{D-1}}{\rho_H^{d-2}}\left(\int_0^{1-\delta}\frac{d\xi}{\sqrt{V_{eff}(\rho_t\xi)}}+
\int_{\frac{\epsilon}{\rho_H}}^1\frac{(1-\xi^{2(d-1)})d\xi}{\xi^{2(d-1)}{\sqrt{V_{eff}(\rho_t\xi)}}}\right).\ee
The second term diverges at the UV limit and it can be regularized
as \eqref{regu} then one can write \be {\cal
A}\approx\frac{L^{D-1}}{\rho_H^{d-2}}\left(\frac{\ell}{2\rho_H}+\frac{1}{(d-2)\epsilon^{d-2}}-c_2\right),\ee
thus the saturated entanglement entropy can be written
accordingly.


\end{document}